\documentclass[12pt,epsf]{article}
\usepackage{graphicx}
\usepackage{epsfig}
\usepackage{latexsym}
\begin{document}

\def\a{\alpha}
\def\b{\beta}
\def\c{\varepsilon}
\def\d{\delta}
\def\e{\epsilon}
\def\f{\phi}
\def\g{\gamma}
\def\h{\theta}
\def\k{\kappa}
\def\l{\lambda}
\def\m{\mu}
\def\n{\nu}
\def\p{\psi}
\def\q{\partial}
\def\r{\rho}
\def\s{\sigma}
\def\t{\tau}
\def\u{\upsilon}
\def\v{\varphi}
\def\w{\omega}
\def\x{\xi}
\def\y{\eta}
\def\z{\zeta}
\def\D{\Delta}
\def\G{\Gamma}
\def\H{\Theta}
\def\L{\Lambda}
\def\F{\Phi}
\def\P{\Psi}
\def\S{\Sigma}

\def\o{\over}
\def\beq{\begin{eqnarray}}
\def\eeq{\end{eqnarray}}
\def\lsim{\mathrel{\rlap{\lower4pt\hbox{\hskip1pt$\sim$}}
    \raise1pt\hbox{$<$}}}              
\def\gsim{\mathrel{\rlap{\lower4pt\hbox{\hskip1pt$\sim$}}
    \raise1pt\hbox{$>$}}}               
\newcommand{\vev}[1]{ \left\langle {#1} \right\rangle }
\newcommand{\bra}[1]{ \langle {#1} | }
\newcommand{\ket}[1]{ | {#1} \rangle }
\newcommand{\EV}{ {\rm eV} }
\newcommand{\KEV}{ {\rm keV} }
\newcommand{\MEV}{ {\rm MeV} }
\newcommand{\GEV}{ {\rm GeV} }
\newcommand{\TEV}{ {\rm TeV} }

\def\diag{\mathop{\rm diag}\nolimits}
\def\Spin{\mathop{\rm Spin}}
\def\SO{\mathop{\rm SO}}
\def\O{\mathop{\rm O}}
\def\SU{\mathop{\rm SU}}
\def\U{\mathop{\rm U}}
\def\Sp{\mathop{\rm Sp}}
\def\SL{\mathop{\rm SL}}
\def\tr{\mathop{\rm tr}}

\def\IJMP{Int.~J.~Mod.~Phys. }
\def\MPL{Mod.~Phys.~Lett. }
\def\NP{Nucl.~Phys. }
\def\PL{Phys.~Lett. }
\def\PR{Phys.~Rev. }
\def\PRL{Phys.~Rev.~Lett. }
\def\PTP{Prog.~Theor.~Phys. }
\def\ZP{Z.~Phys. }

\newcommand{\non}{\nonumber}
\newcommand{\bea}{\begin{eqnarray}}  \newcommand{\eea}{\end{eqnarray}}
\newcommand{\la}{\left\langle} \newcommand{\ra}{\right\rangle}
\def\lrf#1#2{ \left(\frac{#1}{#2}\right)}
\def\lrfp#1#2#3{ \left(\frac{#1}{#2}\right)^{#3}}


\baselineskip 0.7cm

\begin{titlepage}

\begin{flushright}
UT-09-30\\
IPMU 09-0163
\end{flushright}

\vskip 1.35cm
\begin{center}
{\large \bf
Phenomenological Aspects of
Gauge Mediation with Sequestered Supersymmetry Breaking \\ 
in light of Dark Matter Detection
}
\vskip 1.2cm
Motoi Endo, Satoshi Shirai, and Kazuya Yonekura
\vskip 0.4cm

{\it Department of Physics, University of Tokyo,\\
     Tokyo 113-0033, Japan}
\vskip 0.1in
{\em Institute for the Physics and Mathematics of the Universe,\\
University of Tokyo, Chiba 277-8568, Japan}

\vskip 1.5cm

\abstract{ 
Recently, a model of gauge mediation with 
sequestered supersymmetry (SUSY) breaking was proposed.
In this model, the mass of the gravitino is  ${\cal O}(100)~\GEV$ 
without causing the flavor-changing neutral-current problem.
In contrast to traditional gauge mediation, the gravitino is not the 
lightest SUSY particle, and the neutralino is the candidate of the 
dark matter. 
In this paper, we investigate phenomenological aspects of the model 
and discuss the possibility of the direct detection of the dark matter. 
In particular, we focus on the light neutralino case and find that the 
light-Higgsino scenario such as the focus point is interesting 
in the light of the recent CDMS result.
 }
\end{center}
\end{titlepage}

\setcounter{page}{2}

\section{Introduction}
The origin of the dark matter (DM) is one of the most challenging problems
in particle physics and cosmology.
In the framework of the minimal supersymmetric (SUSY) standard model (MSSM),
the lightest SUSY particle (LSP) is a candidate of the DM, since it is stable due to
the R-parity conservation.
In the MSSM, the gravitino or the lightest neutralino, which is the mixed state of the 
Bino, neutral Wino and Higgsino, is appropriate for the DM candidate.
Among them, the Bino-like neutralino DM is attractive.
Its mass is predicted as $m_{\rm DM}\simeq {\cal O}(10-100)$ GeV to explain the 
present DM abundance.
This mass scale is favorable for the discovery of the SUSY particles at the LHC.
In addition, the Bino tends to be naturally lighter than the other SUSY particles 
because the $\U(1)_Y$ gauge interaction is weak.

In the framework of the gravity mediation scenario, the Bino DM can be naturally realized. 
However, the gravity mediation generally suffers from the serious flavor-changing 
neutral-current (FCNC) problem.
Such a problem gives arise due to the flavor-violating non-renormalizable operators
at the Planck scale.
On the other hand, the gauge mediated SUSY breaking model (GMSB) \cite{GM} with 
$m_{3/2}\lsim 1$ GeV is free from the FCNC problems. 
Then, the gravitino is the LSP, and the neutralino cannot be the DM.

In a recent work \cite{Shirai:2008qt}, two of the authors with F.Takahashi and T.T.Yanagida have shown
that if the conformal sequestering occurs in the SUSY breaking sector,
gauge mediation with $m_{3/2} = {\cal O}(100)~\GEV$ can be realized without conflicting 
with the FCNC problem (see also Ref.~\cite{Craig:2008vs}).
In this case, the Bino-like neutralino is the candidate of the DM.
In this paper,  
we investigate the neutralino DM in GMSB model and discuss the detection possibility of 
the DM in the current and future experiments.

\section{Gauge mediation with sequestered SUSY breaking}

For the neutralino to be the LSP, the gravitino must be heavier than the neutralino, i.e. 
$m_{{\tilde \chi}^0_1}<m_{3/2}$.
Then gravity mediation effects are generally non-negligible, which lead to the dangerous FCNC.
The effects come from Planck suppressed operators in a K\"ahler potential of the form
\beq
K \supset \sum_{i,j}\frac{C_{ij}}{M_{PL}^2} S^\dagger S \f^\dagger_i \f_j, \label{eq:sugraeffect}
\eeq
where $S$ is a SUSY breaking chiral superfield in a hidden sector with $\vev{S}=F\h^2$, 
$\f_i$ the MSSM matter fields, $M_{PL} \simeq 2.4\times 10^{18}~\GEV$ the reduced Planck mass,
and $C_{ij}$ some unknown constants which are supposed to be ${\cal O}(1)$. Here $i,j$ are flavor indices.
If $C_{ij}$ is non-diagonal, those operators give a flavor-dependent 
mass matrix to the sfermions of order $m_{3/2}$.

\begin{figure}[t!]
\begin{center}
\scalebox{1}{
\begin{picture}(200,200)(0,-10)
\put(100,216){Energy scale }
\put(100,-10){\vector(0,1){220}}
\put(100,0){\line(1,0){10}}
\put(112,-4){\small electroweak breaking}
\put(100,70){\line(1,0){10}}
\put(112,66){$M_{\rm mess}$}
\put(118,52){$\to$ $m_{\rm soft}$ {\small generation}}
\put(100,110){\line(1,0){10}}
\put(112,106){$\Lambda_{ \scalebox{4}[0.8]{/}\hspace{-2em}{\rm SUSY}}=\sqrt{F}={\cal O}(10^{10})$ GeV}
\put(100,170){\line(1,0){10}}
\put(112,166){\small conformal fixed point $M_*$}
\put(112,130){\scalebox{1.5}[3.8]{\}}}
\put(126,145){\small conformal sequestering}
\put(130,135){\small $\to$ suppression of flavor violating terms}
\put(100,190){\line(1,0){10}}
\put(112,186){$\Lambda_{\rm cut-off }$ such as $M_{ PL}$}

\put(90,23){\scalebox{1.5}[4.6]{\{}}
\put(25,31){\small MSSM RGE}
\end{picture}
}
\caption{Schematic of our model
 }
\label{fig:massscale}
\end{center}
\end{figure}
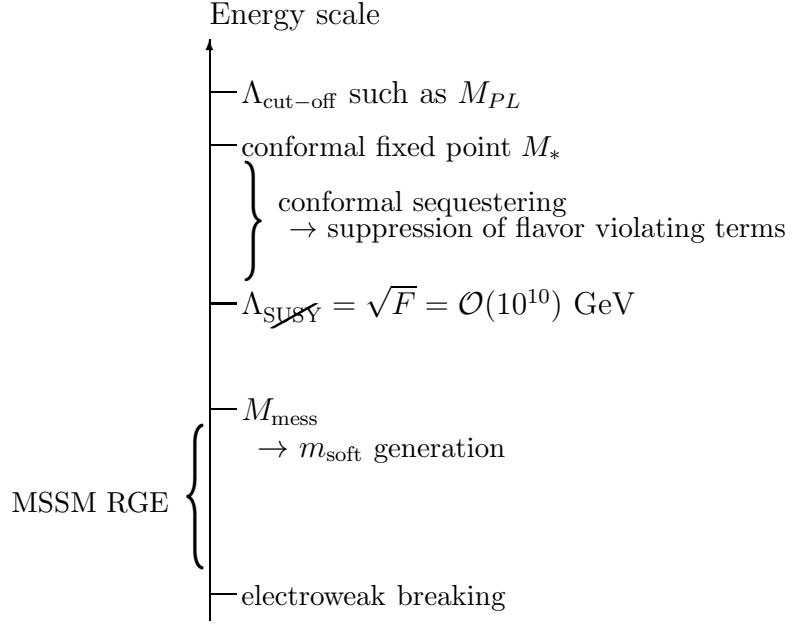

In this section, we review the work of Ref.~\cite{Shirai:2008qt}. There, it was discussed that the neutralino becomes a possible 
candidate of the LSP in gauge mediation~\footnote{
There are also hybrid models of gauge and gravity mediation with the neutralino DM. See e.g., Refs.~\cite{Dudas:2008eq,Dudas:2008qf}.
}, by suppressing the operators (\ref{eq:sugraeffect})
using the conformal sequestering mechanism~\cite{Luty:2001jh,Dine:2004dv,Ibe:2005pj,Schmaltz:2006qs,Kachru:2007xp,Murayama:2007ge}. 
If the SUSY breaking hidden sector is near a conformal fixed point
above the SUSY breaking scale,
renormalization group (RG) effects make the operators (\ref{eq:sugraeffect}) to become~\footnote{
We neglect operator mixings for simplicity.}
\beq
\sum_{i,j}\frac{C_{ij}}{M_{PL}^2} S^\dagger S \f^\dagger_i \f_j \to \left(\frac{\mu_R}{M_*}\right)^{b}\sum_{i,j}\frac{C_{ij}}{M_{PL}^2} S^\dagger S \f^\dagger_i \f_j,
\eeq
where $\mu_R$ is a renormalization scale, $M_{*}$ the scale at which the hidden sector flows near the conformal fixed point. The constant $b$ 
is the minimum eigenvalue of the matrix $(\partial \b_i/\partial g_j)$ evaluated at the fixed point, where $g_i$ are coupling constants and $\b_i$ the beta functions of $g_i$. If the fixed point is infrared stable, we have $b>0$. Therefore, the operators Eq.~(\ref{eq:sugraeffect}) are suppressed.
This suppression continues until the conformal invariance breaks down, and the scale of breakdown is almost equal to the SUSY breaking scale 
$\mu_R\sim\sqrt{|F|}$ in the model of Refs.~\cite{Ibe:2005pj,Shirai:2008qt}.
In Fig.~\ref{fig:massscale}, we show the schematic of the model.
After the suppression, the gravity mediation effects give a mass matrix to the sfermions of order
\beq
m^2_{\rm grav} &\sim& C\left(\frac{\sqrt{M_{PL}m_{3/2}}}{M_*}\right)^{b}m^2_{3/2} \nonumber \\ &\sim&  
(100~\GEV)^2 \times C \left(10^{-8}\cdot \sqrt{\frac{m_{3/2}}{100~\GEV}}\cdot\frac{10^{18}~\GEV}{M_*} \right)^b \left(\frac{m_{3/2}}{100~\GEV}\right)^2,
\eeq
where $C$ collectively denotes $C_{ij}$, and we have used $m_{3/2}=|F|/\sqrt{3}M_{PL}$.
If $M_{*}$ is sufficiently large, and $b$ is $O(1)$ (i.e., the fixed point is strongly coupled), we can suppress the gravity mediation contributions. 
Assuming that $C\sim {\cal O}(1)$ and $m_{3/2}\sim {\cal O}(100)~\GEV$, 
we require~\cite{Gabbiani:1996hi}
\beq
\e \equiv \left(10^{-8}\cdot \sqrt{\frac{m_{3/2}}{100~\GEV}}\cdot\frac{10^{18}~\GEV}{M_*} \right)^b \lsim 10^{-4},
\eeq
to avoid the FCNC.

Gravity mediation contribution to the A-term is also suppressed. The A-term is generated, e.g., by the operator~\footnote{The A-term is also produced by
operators of the form $S^\dagger \f^\dagger \f /M_{PL}+{\rm h.c.}$ in the K\"ahler potential, but the argument below applies with almost no change.}
\beq
\int d^2 \h \sum \frac{C'_{ijk}}{M_{PL}}S\f_i\f_j\f_k.
\eeq
This term is suppressed to
\beq
\ \left(\frac{\sqrt{|F|}}{M_*}\right)^{\g_S/2} \int d^2\h \sum \frac{C'_{ijk}}{M_{PL}}S\f_i\f_j\f_k ,
\eeq
where $\g_S$ is the anomalous dimension of $S$ at the conformal fixed point. This gives the A-term of order
\beq
a_{ijk} \simeq 100~\GEV \times C'_{ijk} \left(10^{-8}\cdot \sqrt{\frac{m_{3/2}}{100~\GEV}}\cdot\frac{10^{18}~\GEV}{M_*} \right)^{\g_S/2} \left(\frac{m_{3/2}}{100~\GEV}\right).
\eeq
It is known that $\g_S$ is positive from the unitarity bound of conformal field theory, and $\g_S$ can be as large as $\g_S=2$. Then, the suppression factor 
becomes as large as ${\cal O}(10^{-8})$ if $\g_S \simeq 2$ and $M_{*} \sim 10^{18}~\GEV$.
Thus, this term becomes negligible even if $C'_{ijk} \sim {\cal O}(1)$ (i.e. even without assuming that $C'_{ijk}$ is proportional to the MSSM Yukawa couplings.)

There is another important effect in the conformal fixed point dynamics~\cite{Shirai:2008qt}. 
Let us consider a gauge mediation model with a superpotential
\beq
W=M\bar{\P}\P+y S \bar{\P}\P,
\eeq
where $\P$ and $\tilde{\P}$ are messenger fields. Due to the RG effect, the Yukawa coupling 
constant $y$ becomes suppressed. 
Assuming that the messenger mass scale is smaller than the SUSY breaking scale $\sqrt{|F|}$, the Yukawa coupling at the 
messenger scale $y_{\rm mess}$ is related to the one at the scale $M_*$, $y_{0}$, as follows:
\beq
y_{\rm mess} &\simeq& y_0 \left(\frac{\sqrt{|F|}}{M_*}\right)^{\g_S/2}  = y_0 \left(10^{-8}\cdot \sqrt{\frac{m_{3/2}}{100~\GEV}}\cdot\frac{10^{18}~\GEV}{M_*} 
\right)^{\g_S/2} \nonumber \\
&=&y_0 \e^{\g_S \o 2b}. \label{eq:smallyukawa}
\eeq 
Thus, the Yukawa coupling constant is naturally small. Actually, it is suppressed by $\e^{\g_S / 2b}$, 
at the messenger mass  scale. This leads to a small messenger mass scale if
we fix $\Lambda =yF/M$ and $m_{3/2}=F/\sqrt{3}M_{PL}$.

Lastly, let us touch on the $\mu$ problem in the model. In the gauge mediation, the problem is 
generally severe. However, it may be solved when the gravitino mass is 
$m_{3/2} \sim {\cal O}(100)~\GEV$ \cite{Shirai:2008qt}.

\section{Dark matter and MSSM mass spectrum}

Recently, CDMS collaboration reports two candidate events for the DM scatterings \cite{Ahmed:2009zw}.
Although this number is too low to confirm the DM detection, compared with the expected background event rate, 
they might be the first signal of the DM. Since the events have relatively low recoil energy, $11$ and $15~\KEV$, 
a light DM may be preferred, i.e. the DM mass does not far deviate from 100~GeV. 
Thus, we explore two cases in this section: the neutralino mass is above 100~GeV and 
smaller than it.
At first, we consider the neutralino mass above 100~GeV. Here, we take account of the other phenomenological constraints, e.g. the LEP bounds 
on the particle masses and those from $b \to s \gamma$ and muon $g-2$. In the following analysis, 
we take the sign of the $\mu$ parameter to be positive. Secondary, we consider a lighter neutralino scenario, 
$m_{\tilde {\chi}^0_1} < 100$~GeV. This case may be particularly interesting, since the direct detection experiments are 
sensitive for light DM of the mass $30-60$ GeV. Thus, we will study the detection possibility of the light neutralino 
DM with the direct detection experiments.

\subsection{Heavy neutralino}
Firstly, let us discuss the heavy neutralino DM ($m_{\rm DM} \gsim 100$ GeV).

\subsubsection*{Minimal GMSB}
We here consider a simple GMSB model, where a SUSY breaking field $S$
couples to a pair of messenger chiral superfields, $\P$ and
$\bar{\P}$, which transform as ${\bf 5}$ and $\bar{\bf 5}$ under the
$\SU(5)_{\rm GUT}$.
The simplest form of the coupling of the messenger and the SUSY breaking field is
\beq
W \;=\; y S\Psi{\bar \Psi}+M \Psi{\bar \Psi},
\eeq
where $M$ is the messenger mass, and $y$ is set to be the value at
the messenger mass scale throughout this section, i.e., $y=y_{\rm mess}$.  
In our model, $y$ is naturally very small (see Eq.~(\ref{eq:smallyukawa})).  
The SUSY breaking chiral field $S$ develops an $F$-term vacuum expectation value (VEV) $\langle S \rangle = \theta^2 F$, which is related to
the gravitino mass as $|F| = \sqrt{3} m_{3/2} M_{PL}$, assuming that
the SUSY breaking is dominated by $F$.

In the GMSB models, the MSSM gaugino masses are generated from loop
diagrams of the messengers. At the one-loop level, the gaugino masses are
given by
\begin{equation}
M_{a} \;=\; \frac{\alpha_a}{4\pi}\Lambda_{eff} (1+{\cal O}(x^2)),
\label{eq:gaugino_mass}
\end{equation}
where we have defined $\Lambda_{eff} =y F/M$ and $x=y F/M^2$.
Here $a = 1,2, 3$ labels U(1), SU(2) and SU(3) in the MSSM, respectively, 
and we use the normalization $\alpha_1=5 \alpha _{\rm EM}/(3 \cos^2\theta_{W})$.
The soft scalar masses arise at the  two loop level, and are given by
\begin{equation}
m^2_{\phi_i}\;=\;2\Lambda_{eff} ^2 \sum_a \left(\frac{\alpha_a}{4\pi}\right)^2 C_a (i) (1+{\cal O}(x^2)), \label{eq:scalar_mass}
\end{equation}
where $C_a(i)$ are Casimir invariants for the visible particles $\phi_i$ 
($C_1(i) = 3Y_i^2/5$).
$x$ is bounded as $x<1$ for the messengers not to become tachyonic, and then the corrections of ${\cal O}(x^2)$ are small and we omit these corrections in the following analysis.
We see that $m_{\f_i} \simeq M_a = {\cal O}(1)~\TEV$ is realized for $\L_{eff}={\cal O}(10^5)~\GEV$.

Since the above
expressions for the soft masses are given at the messenger scale, one
should solve the MSSM RG equation to
get the on-shell masses and mixing matrices.

\begin{figure}[h!]
\begin{center}
\epsfig{file=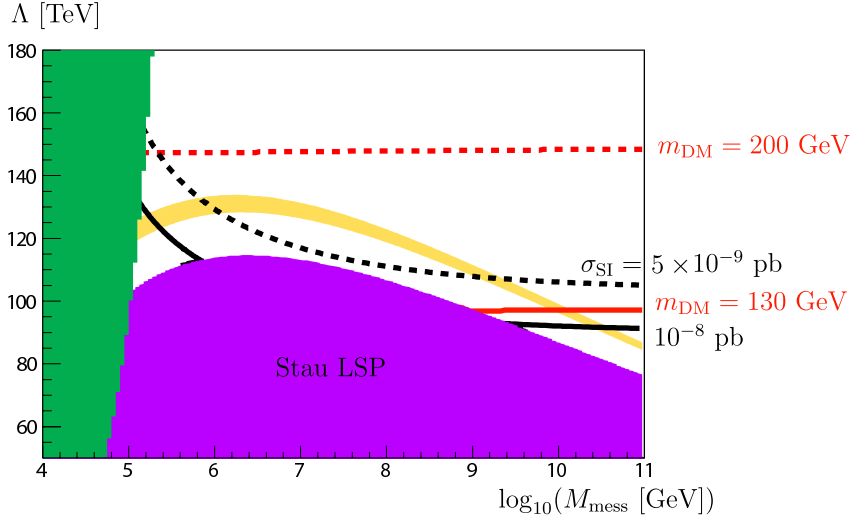,clip,scale=.5}\\
(a)\\
\end{center}
\begin{center}
\epsfig{file=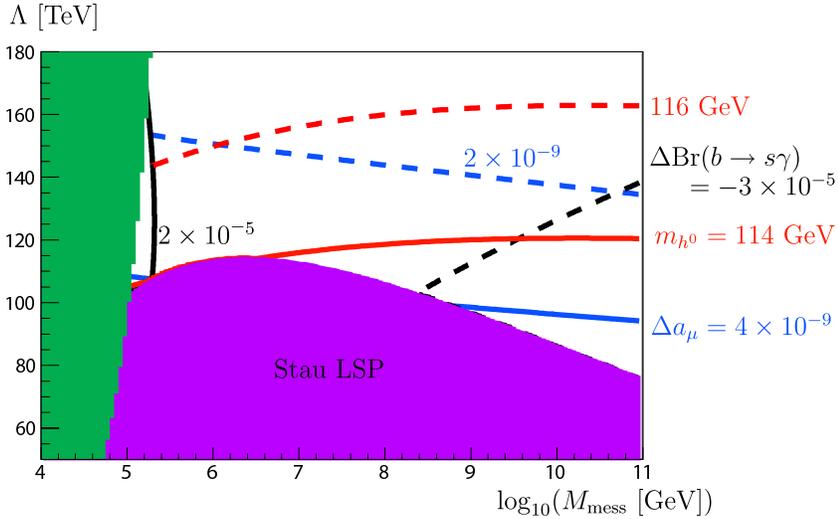,clip,scale=.5}\\
(b)
\end{center}
\caption{(a): Red: DM mass $m_{\rm DM}$, Black: spin independent cross section to a nucleon $\sigma_{\rm SI}$.
The yellow region shows $0.08<\Omega h^2<0.12$.
(b): Red: Higgs mass $m_{h^0}$, Blue: $\Delta a_{\mu}$, Black $\Delta {\rm Br}(b\to s \gamma)$.
In the green region, the messengers are tachyonic and in the purple region, the stau is the LSP.
 }
\label{fig:heavyDM}
\end{figure}

In Fig.~\ref{fig:heavyDM}-(a), we show the DM mass $m_{\rm DM}$, relic abundance $\Omega h^2$ and spin-independent cross section to 
a nucleon $\s_{\rm SI}$, and in Fig.~\ref{fig:heavyDM}-(b), we show the Higgs mass $m_{h^0}$, the difference of the muon anomalous magnetic moment
$\Delta a_{\mu} \equiv a_{\mu}|_{\rm MSSM}-a_{\mu}|_{\rm SM}$, and the difference of the branching fraction of $b \to s\gamma$, 
$\Delta{\rm Br}(b\to s\gamma ) \equiv {\rm Br}(b\to s\gamma )|_{\rm MSSM}- {\rm Br}(b\to s\gamma )|_{\rm SM}$.
Here, we set $\tan\beta=40$.
To calculate the MSSM mass spectrum and the DM property, we have
used the programs \verb+SOFTSUSY+ 2.0.18~\cite{Allanach:2001kg} and
\verb+micrOMEGAs+ 2.2~\cite{micromegas}.
$\Delta{\rm Br}(b\to s\gamma ) $ is calculated
with \verb+SusyBSG+ 1.3.1 \cite{Degrassi:2007kj}.
As pointed out in Ref.~\cite{Shirai:2008qt}, when $\tan\beta$ is large, the stau mass becomes smaller, and thus the coannihilation 
effect is essential for the correct DM abundance.
Therefore the masses of the DM and sleptons are degenerate.

\subsection{Light neutralino}

In this subsection, we study the case of the light DM,
paying particular attention to the experimental constraints from collider and precision measurements.

\subsubsection*{Minimal GMSB}
\begin{figure}[h!]
\begin{tabular}{cc}
\begin{minipage}{0.5\hsize}
\begin{center}
\epsfig{file=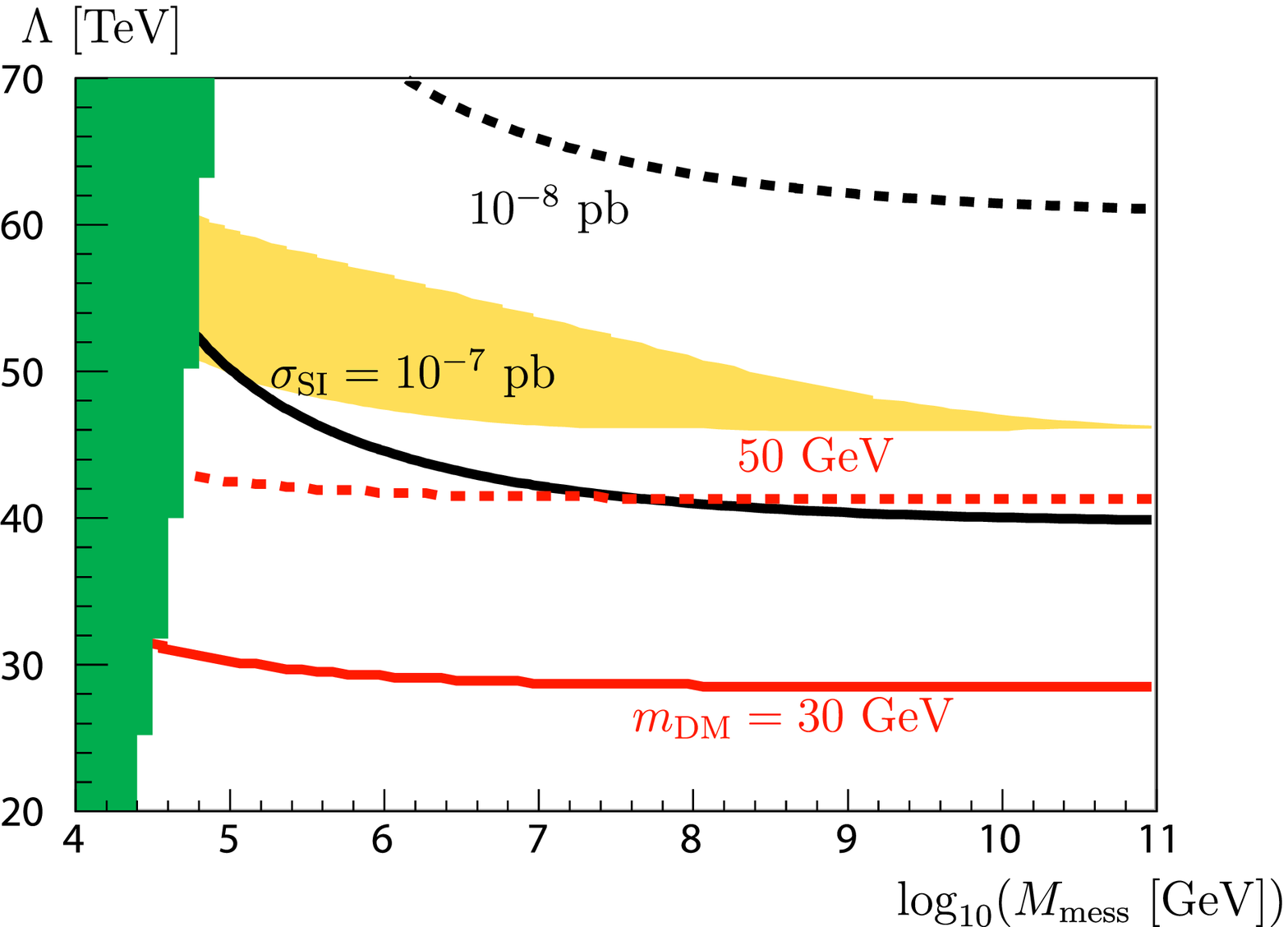,clip,scale=.46}
(a)
\end{center}
\end{minipage}
\begin{minipage}{0.5\hsize}
\begin{center}
\epsfig{file=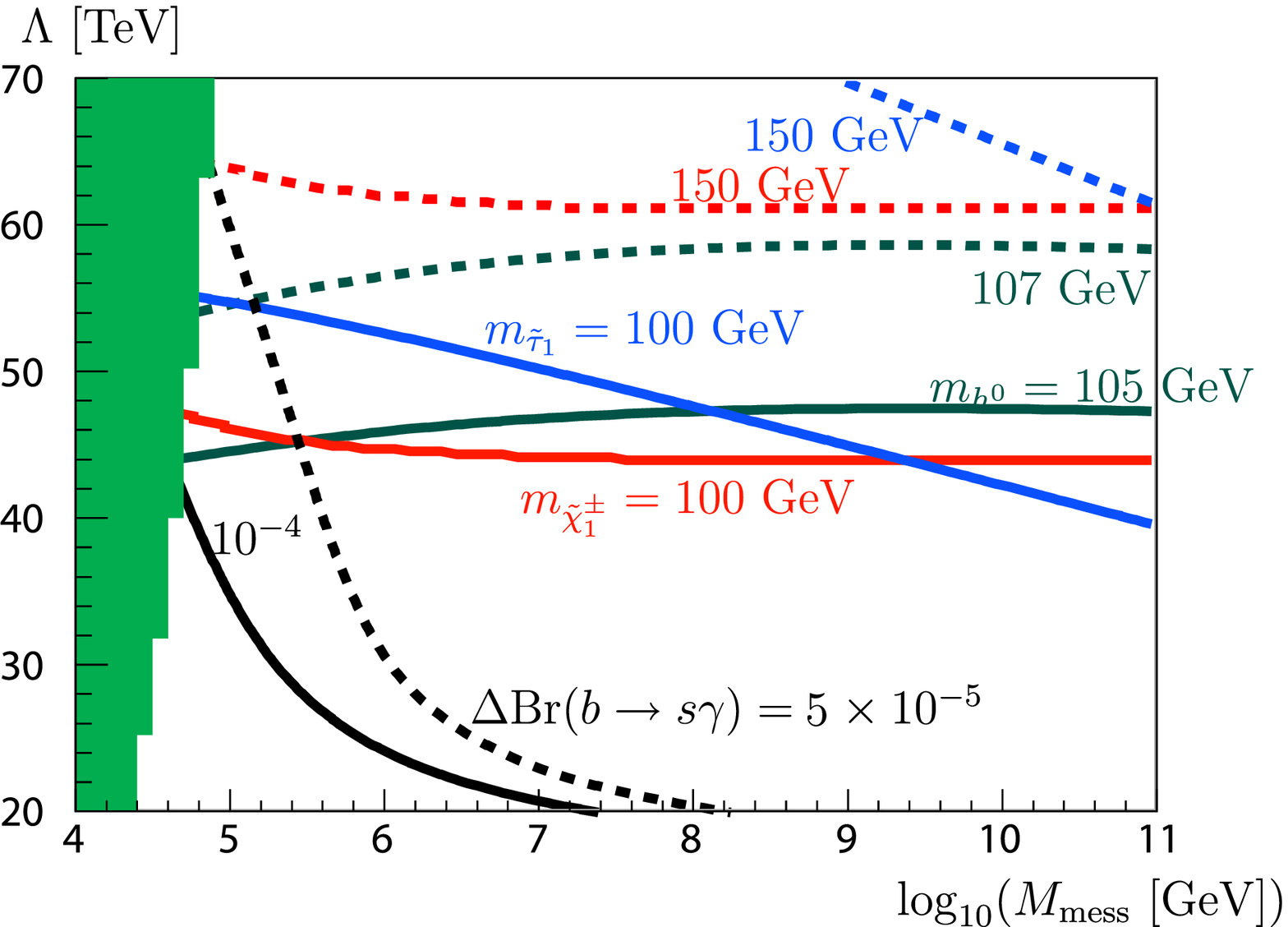,clip,scale=.46}
(b)
\end{center}
\end{minipage}
\end{tabular}
\caption{Some physical quantities in mGMSB with $\tan\beta=10$.
(a): red: DM mass, black: cross spin-independent section to a nucleon and yellow region shows $0.08<\Omega h^2<0.12$.
(b): red: $m_{\tilde{\chi}^{\pm}_1}$, black: $\Delta{\rm Br}(b\to s\gamma )$, green: Higgs mass $m_{h^0}$, blue: $m_{\tilde{\tau}_1}$.
In the green region, the messengers are tachyonic. 
 }
\label{fig:DMN-10}
\end{figure}

\begin{figure}[htbp]
\begin{tabular}{cc}
\begin{minipage}{0.5\hsize}
\begin{center}
\epsfig{file=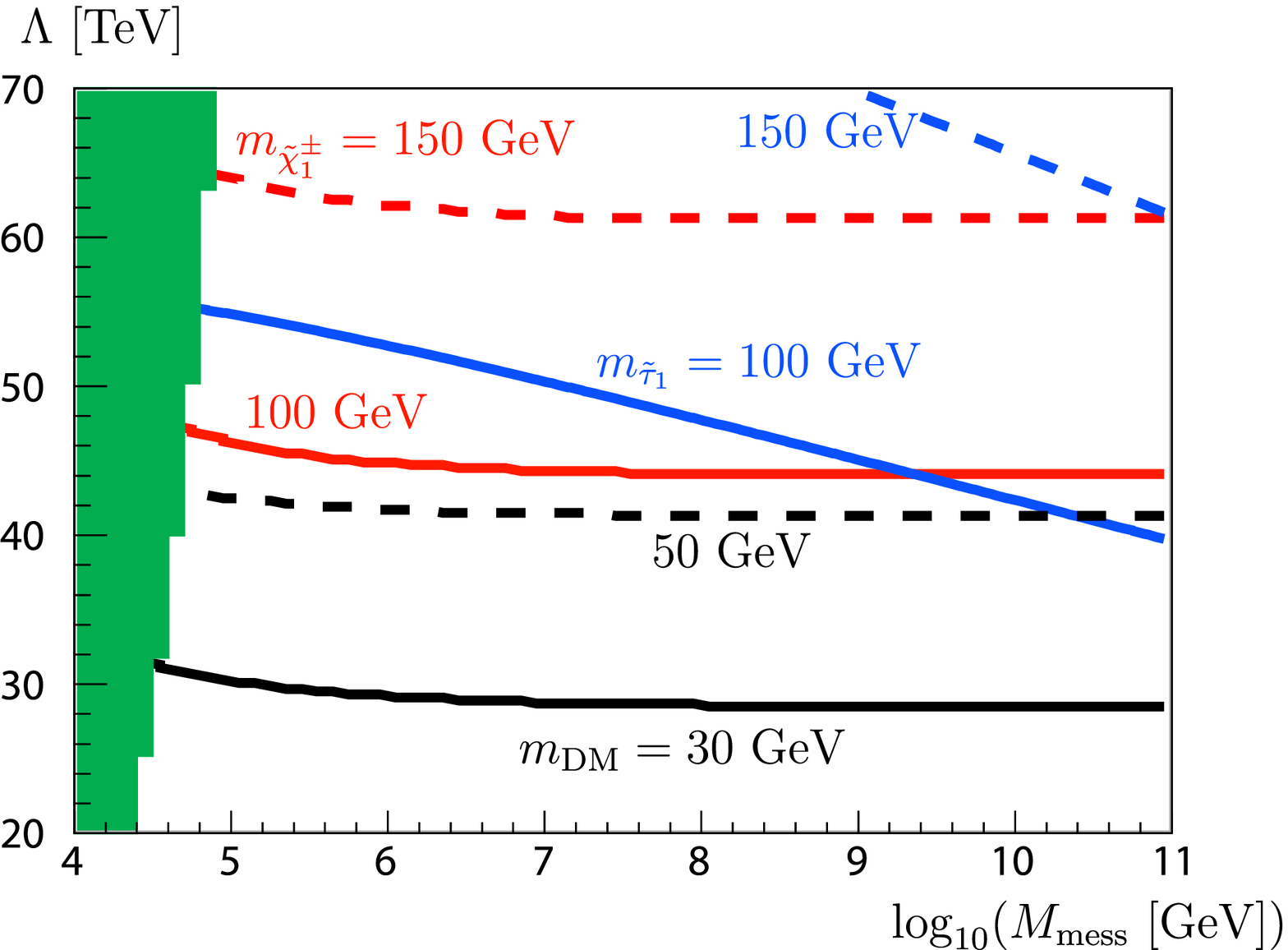,clip,scale=.46}
(a) ${\bf 5}+\bar{\bf 5}$
\end{center}
\end{minipage}
\begin{minipage}{0.5\hsize}
\begin{center}
\epsfig{file=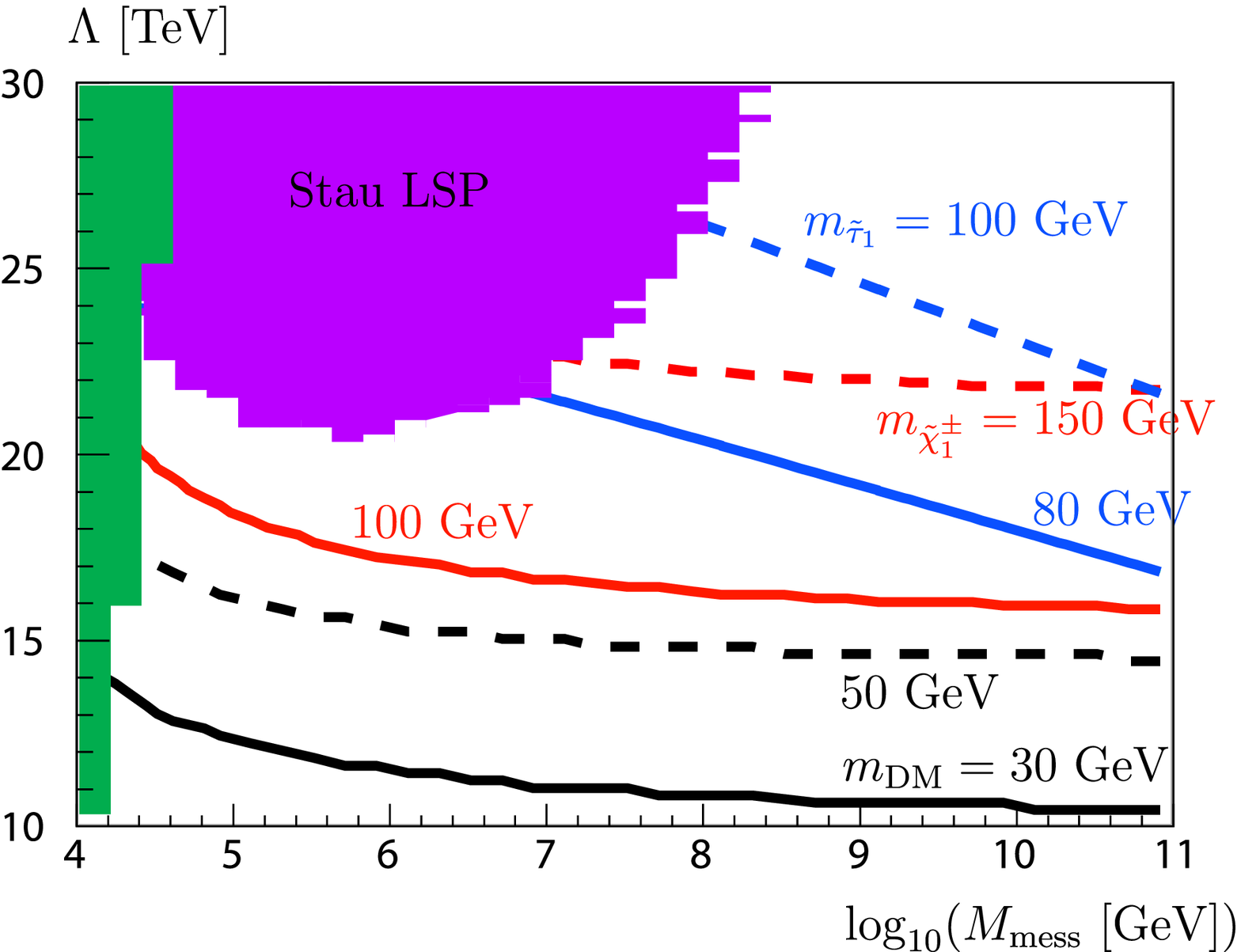,clip,scale=.46}
(b) ${\bf 10}+\bar{\bf 10}$
\end{center}
\end{minipage}
\end{tabular}
\caption{Masses of the lightest neutralino, chargino and stau.
 }
\label{fig:MASS}
\end{figure}

In Figs.~\ref{fig:DMN-10}, we show some parameters of the DM and MSSM spectrum with $\tan\beta=10$.
In the minimal GMSB, the Bino-like  neutralino can be the LSP and the DM.
However, if the mass of Bino-like  neutralino is lighter than about 50 GeV, 
the lightest chargino is likely to be lighter than about $100~\GEV$, which conflicts with the current collider experiments.
In Figs.~\ref{fig:MASS}, we show the masses of the lightest neutralino, chargino and stau.
We also studied the case that the representation of the messenger is  ${\bf 10}+\bar{\bf 10}$.
Here, we set $\tan\beta=10$.
One can see that, if the mass of the neutralino is $(30-50)~\GEV$,
some charged SUSY particles are  too light to evade the LEP bound.
In general, if the messenger belongs to a higher dimensional representation such as ${\bf 24}$, 
the scalar particles tend to be lighter, and the situation gets worse.

\subsubsection*{Non-minimal GMSB }
Alternatively, we modify the coupling between the messenger and the SUSY breaking field as
\beq
W \;=\;  \frac{\lambda}{M_P}S\Psi\langle{\bf 24}\rangle{\bar \Psi}+M \Psi{\bar \Psi},
\eeq
where $\langle{\bf 24}\rangle$ is the VEV of an SU(5)$_{\rm GUT}$ adjoint field, which may be the GUT breaking Higgs field (a similar idea was used in
Ref.~\cite{Dudas:2008eq}).
By inserting $\langle{\bf 24}\rangle = v\cdot{\rm diag}(3,3,-2,-2,-2)$, we have
\beq
W \;=\; y_{\ell} S \Psi_{\ell}{\bar \Psi}_{\ell} - \frac{2}{3}y_{\ell} S \Psi_{d}{\bar \Psi}_{d} +
M (\Psi_d{\bar \Psi}_d + \Psi_{\ell}{\bar \Psi}_{\ell}),
\eeq
where $y_{\ell} =3\l v /M_{P}$.
In this case, the down-type  $\P_{d}$ and lepton-type $\P_{\ell}$ messengers  have the different couplings to the SUSY breaking field $S$.

The MSSM gaugino masses are given by
\beq
M_{1} &=& \frac{\alpha_1}{4\pi}\frac{\L_{\ell}}{3},\\
M_{2} &=& \frac{\alpha_2}{4\pi}{\L_{\ell}},\\
M_{3} &=& -\frac{\alpha_3}{4\pi}\frac{2\L_{\ell}}{3},
\eeq
where $\L_{\ell} = y_{\ell} F_S/M$.
The soft scalar masses are given by
\begin{equation}
m^2_{\phi_i}\;=\;2\left(\frac{\alpha_1}{4\pi}\right)^2 C_1 (i) \frac{7\L_{\ell}^2}{9}
+ 2\left(\frac{\alpha_2}{4\pi}\right)^2{C_2 (i)\L_{\ell}^2}
+ 2\left(\frac{\alpha_3}{4\pi}\right)^2C_3 (i) \frac{4\L_{\ell}^2}{9}.\label{eq:scalar_mass2}
\end{equation}
In this case, $m_{\tilde{W}}\simeq 6m_{\tilde{B}}$.
Thus, the lightest chargino can be heavy to evade the experimental bounds, even if
$m_{\rm DM} \simeq 30~\GEV$.

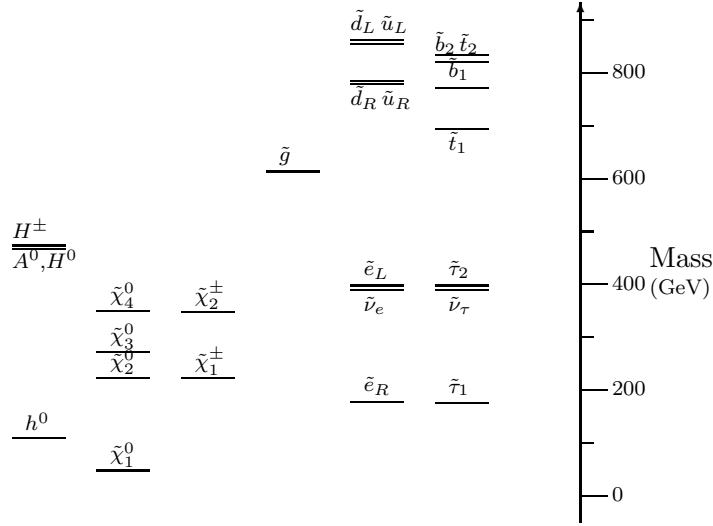
\begin{figure}[t!]
\begin{center}
\scalebox{1}{
\begin{picture}(220,186.881)(0,0)
\put(5,21.8544){\line(1,0){20}}
\put(10,24.8544){$\scriptstyle h^0$}
\put(5,93.2584){\line(1,0){20}}
\put(5,93.3954){\line(1,0){20}}
\put(5,94.7354){\line(1,0){20}}
 \put(5,97.7354){$\scriptstyle H^{\pm}$}
 \put(5,86.2584){$\scriptstyle A^0,H^0$}
\put(37,9.5589){\line(1,0){20}}
\put(42.,13.5589){$\scriptstyle {\tilde{\chi}^0_1}$}
\put(37,44.592){\line(1,0){20}}
\put(42.,48.592){$\scriptstyle {\tilde{\chi}^0_2}$}
\put(37,54.3546){\line(1,0){20}}
\put(42.,58.3546){$\scriptstyle {\tilde{\chi}^0_3}$}
\put(37,69.913){\line(1,0){20}}
\put(42.,73.913){$\scriptstyle {\tilde{\chi}^0_4}$}
\put(69,44.566){\line(1,0){20}}
\put(74.,48.566){$\scriptstyle {\tilde{\chi}^{\pm}}_1$}
\put(69,69.5706){\line(1,0){20}}
\put(74.,73.5706){$\scriptstyle {\tilde{\chi}^{\pm}_2}$}
\put(101,122.757){\line(1,0){20}}
\put(106,126.757){$\scriptstyle {\tilde g}$}
\put(133,172.474){\line(1,0){20}}
\put(133,171.131){\line(1,0){20}}
\put(133,176.474){$\scriptstyle \tilde d_L\,\tilde u_L$}
\put(133,155.789){\line(1,0){20}}
\put(133,156.944){\line(1,0){20}}
\put(133,147.789){$\scriptstyle \tilde d_R\,\tilde u_R$}
\put(133,79.5842){\line(1,0){20}}
\put(133,78.0162){\line(1,0){20}}
\put(138,70.0162){$\scriptstyle \tilde \nu_e$}
\put(138,83.5842){$\scriptstyle \tilde e_L$}
\put(133,35.566){\line(1,0){20}}
\put(138,39.566){$\scriptstyle \tilde e_R $}
\put(165,139.014){\line(1,0){20}}
\put(170,131.014){$\scriptstyle \tilde t_1$}
\put(165,166.881){\line(1,0){20}}
\put(165,154.476){\line(1,0){20}}
\put(170,158.476){$\scriptstyle \tilde b_1$}
\put(165,164.281){\line(1,0){20}}
\put(165,168.281){$\scriptstyle \tilde b_2\,\tilde t_2$}
\put(165,35.0972){\line(1,0){20}}
\put(170,39.0972){$\scriptstyle \tilde \tau_1$}
\put(165,79.5468){\line(1,0){20}}
\put(170,83.5468){$\scriptstyle \tilde \tau_2$}
\put(165,77.9294){\line(1,0){20}}
\put(170,69.9294){$\scriptstyle \tilde \nu_{\tau}$}
\put(246,86.4404){\small Mass }
\put(246,76.4404){$\scriptstyle {\rm (GeV)}$ }
\put(220,-10){\vector(0,1){196.881}}
\put(220,0){\line(1,0){10}}
\put(232,-2){$\scriptstyle 0$}
\put(220,40){\line(1,0){10}}
\put(232,38){$\scriptstyle 200$}
\put(220,80){\line(1,0){10}}
\put(232,78){$\scriptstyle 400$}
\put(220,120){\line(1,0){10}}
\put(232,118){$\scriptstyle 600$}
\put(220,160){\line(1,0){10}}
\put(232,158){$\scriptstyle 800$}
\put(220,0){\line(1,0){5}}
\put(220,20){\line(1,0){5}}
\put(220,40){\line(1,0){5}}
\put(220,60){\line(1,0){5}}
\put(220,80){\line(1,0){5}}
\put(220,100){\line(1,0){5}}
\put(220,120){\line(1,0){5}}
\put(220,140){\line(1,0){5}}
\put(220,180){\line(1,0){5}}
\end{picture} }
\caption{An example of the MSSM mass spectrum in the non-mGMSB model.
 }
\label{fig:mass}
\end{center}
\end{figure}

As an example, in Fig.~\ref{fig:mass} we show the MSSM mass spectrum at the point $M=10^6~\GEV$, $\L_{\ell} = 110~\TEV$, 
 $\tan\beta=10$ and ${\rm sgn}(\mu) = +1$. 

\begin{figure}[htbp]
\begin{tabular}{cc}
\begin{minipage}{0.5\hsize}
\begin{center}
\epsfig{file=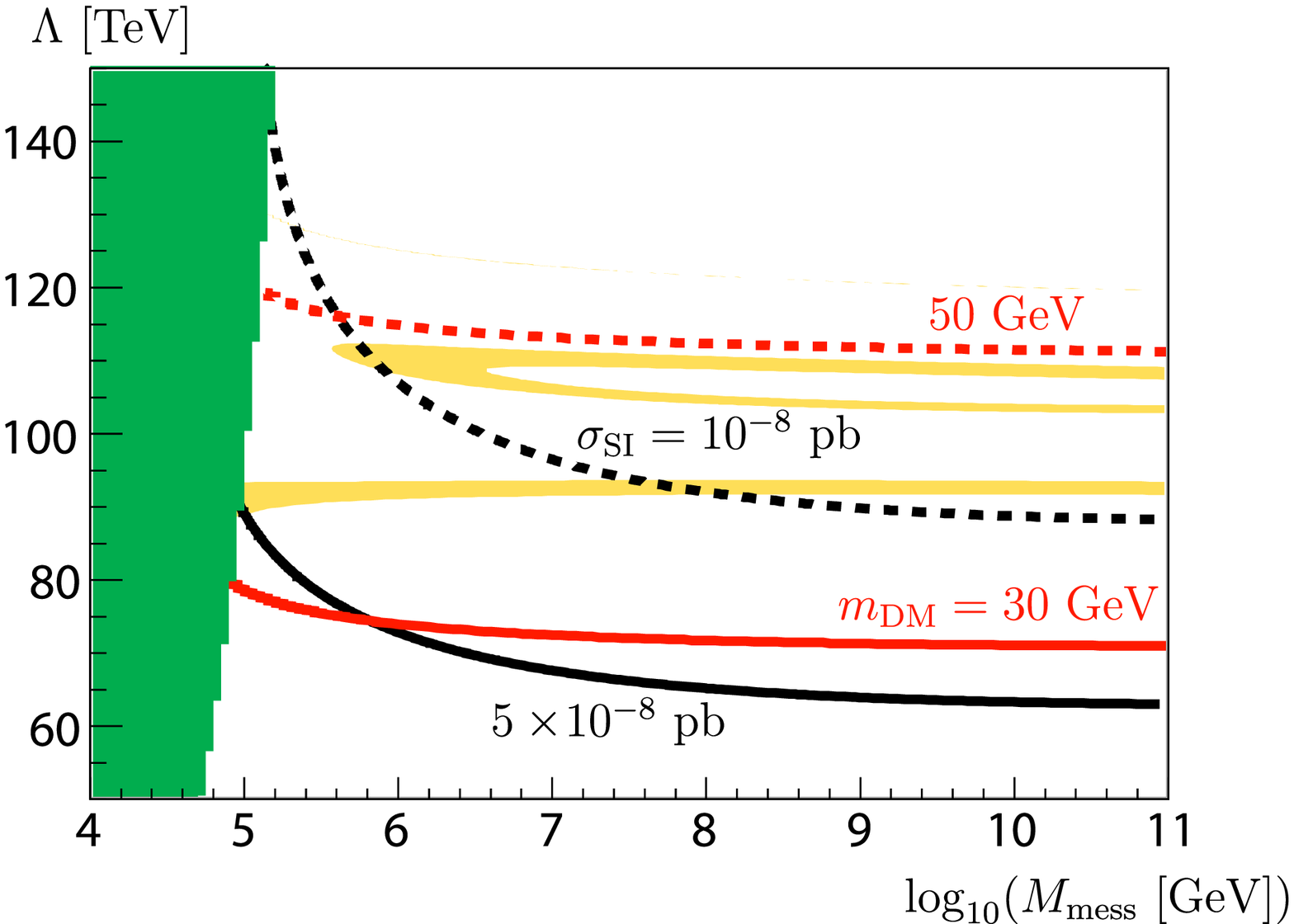,clip,scale=.46}
(a)
\end{center}
\end{minipage}
\begin{minipage}{0.5\hsize}
\begin{center}
\epsfig{file=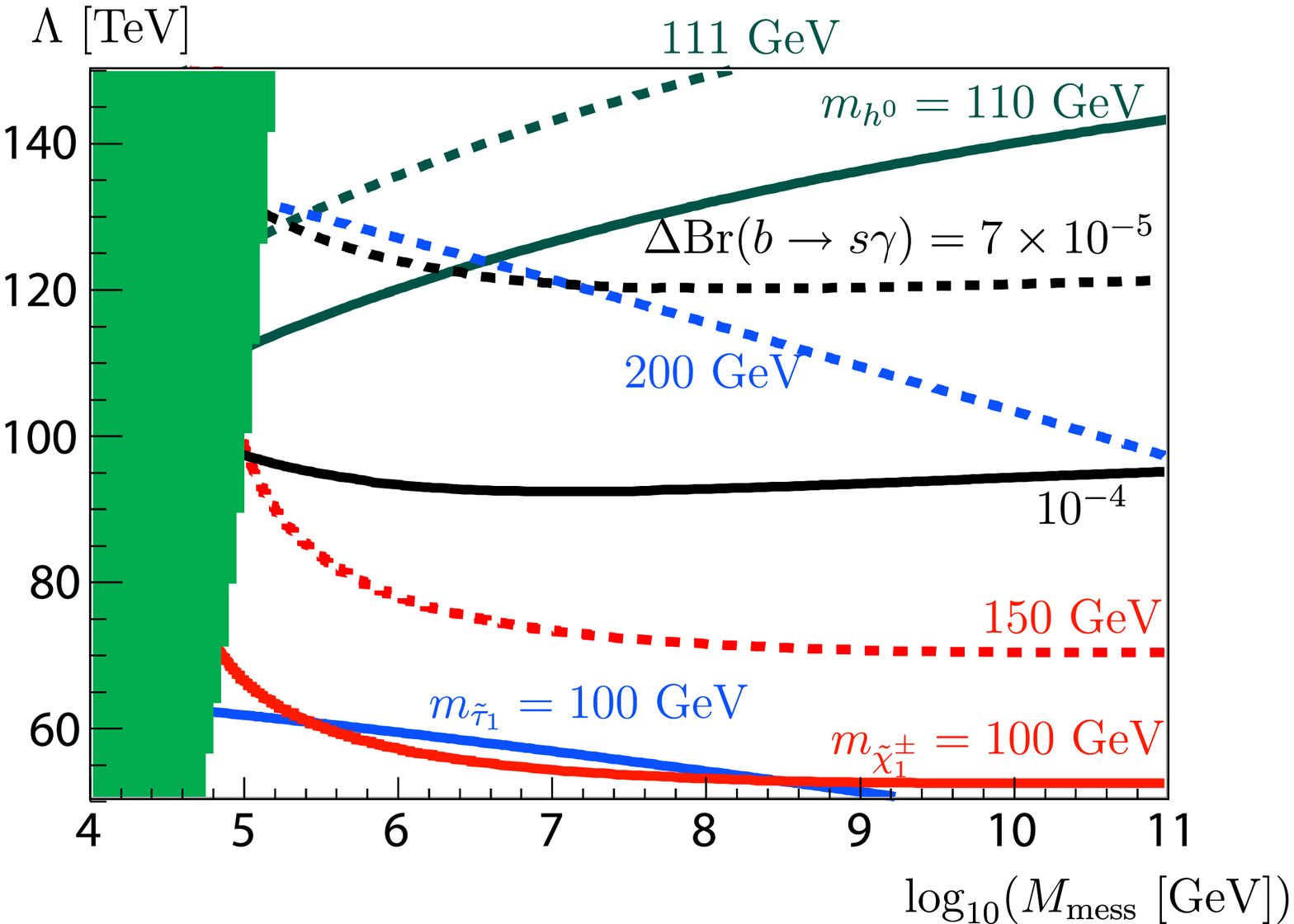,clip,scale=.46}
(b)
\end{center}
\end{minipage}
\end{tabular}
\caption{Same as Figs.~\ref{fig:DMN-10} except non-mGMSB with $\tan\beta=10$.
 }
\label{fig:DM24-10}
\end{figure}

\begin{figure}[htbp]
\begin{tabular}{cc}
\begin{minipage}{0.5\hsize}
\begin{center}
\epsfig{file=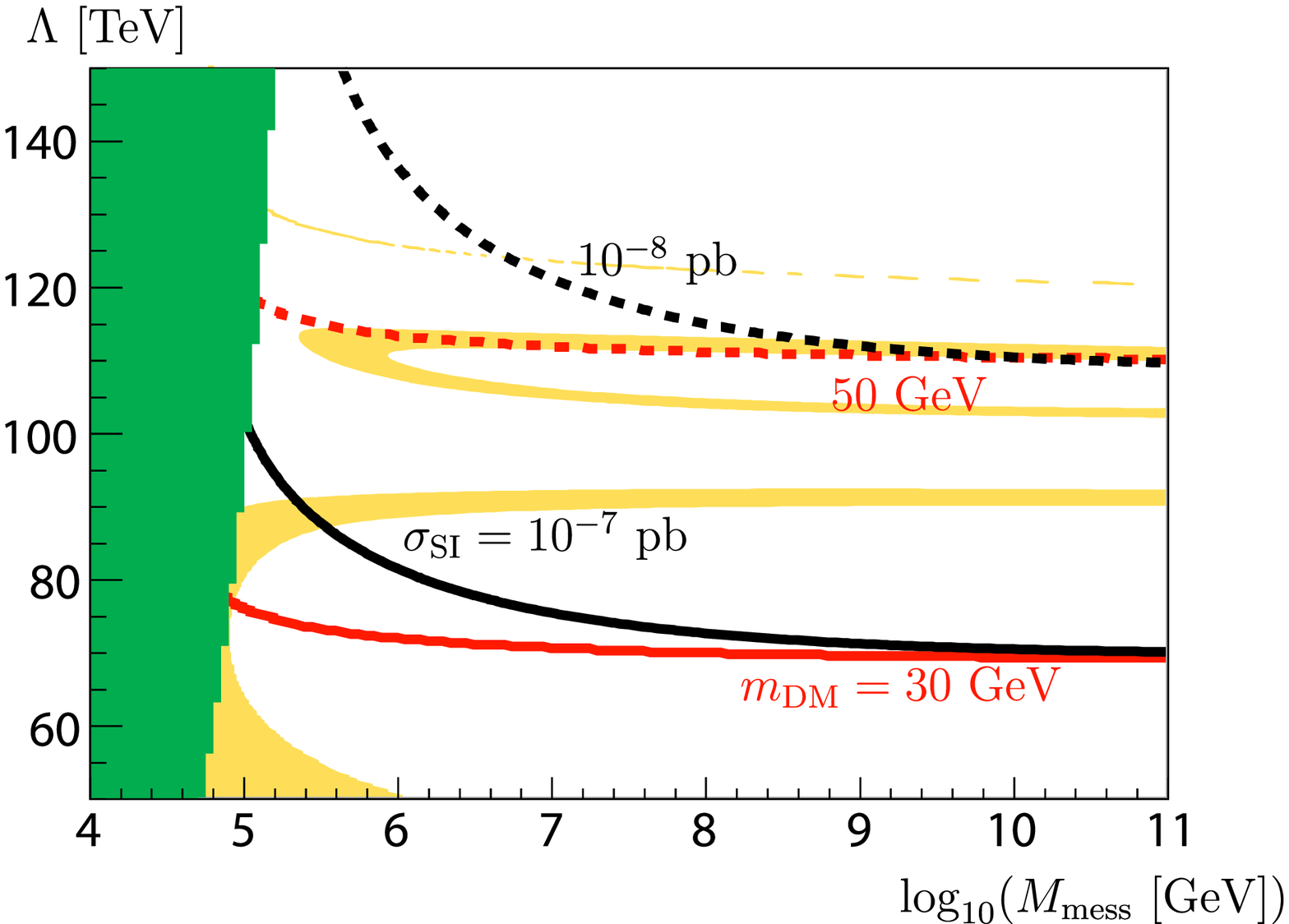,clip,scale=.46}
(a)
\end{center}
\end{minipage}
\begin{minipage}{0.5\hsize}
\begin{center}
\epsfig{file=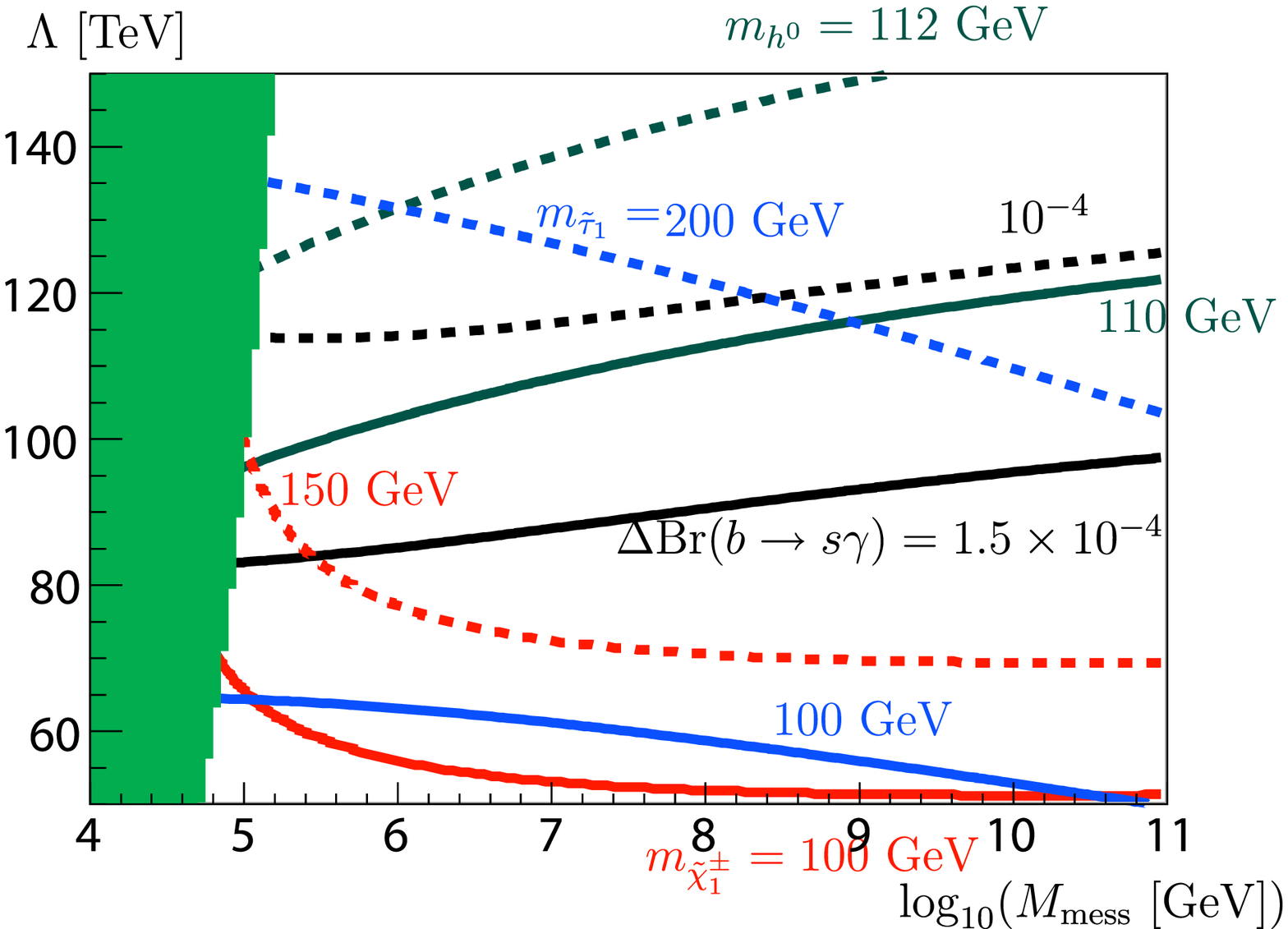,clip,scale=.46}
(b)
\end{center}
\end{minipage}
\end{tabular}
\caption{Same as Figs.~\ref{fig:DMN-10} except non-mGMSB with $\tan\beta=20$.
 }
\label{fig:DM24-20}
\end{figure}

In Figs.~\ref{fig:DM24-10} and \ref{fig:DM24-20},
we show some physical quantities as functions of $\Lambda_{\ell}$ and $M_{\rm mess}$.
We set $\tan\beta=10$ for Figs.~\ref{fig:DM24-10} and $\tan\beta=20$ for Figs.~\ref{fig:DM24-20}.
In this case, the lightest neutralino has sizable components of the Higgsino for a small value of the messenger mass $M$.
There are two reasons for that.
One reason is that the colored SUSY particles are rather light in the non-mGMSB model 
(see Eqs.~(\ref{eq:scalar_mass2}) and (\ref{eq:scalar_mass})). The other reason is that the running of the RG equation is short.

We notice that the sign of the gluino is negative, which is opposite to that of the Wino. This seems to be dangerous, because the 
SUSY contribution to the $b \to s\gamma$ ratio tends to be large. Since the squark masses are around 1~TeV, and the A-parameter 
of the top squark is suppressed, the $b \to s\gamma$ constraint is found to be ameliorated. 

\section{Scan of parameter space}
\begin{figure}[htbp]
\begin{center}
\epsfig{file=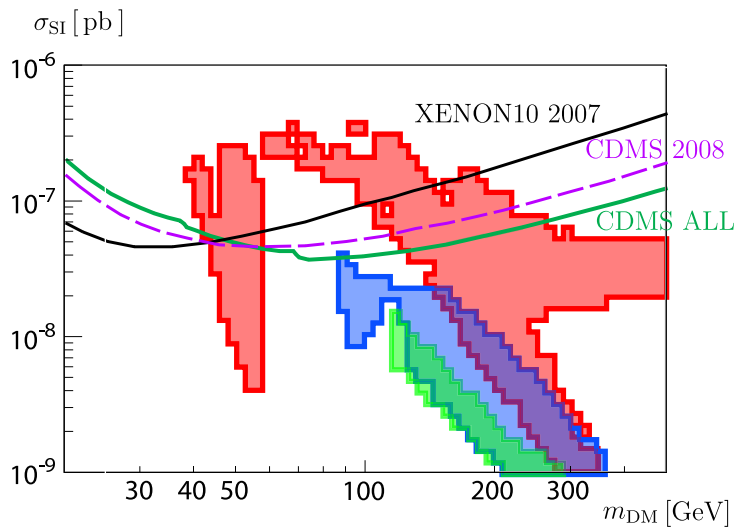,clip,scale=.5}
\end{center}
\caption{DM mass $m_{\rm DM}$ and spin-independent cross section $\s_{\rm SI}$ with experimental data~\cite{Angle:2007uj,Ahmed:2008eu}.
The red region shows non-mGMSB, blue mGMSB with ${\bf 5}+\bar{\bf 5}$ messenger and green mGMSB with  ${\bf 10}+\bar{\bf 10}$ messenger.
 }
\label{fig:result}
\end{figure}

We scan over the parameter space of the model and discuss the detection possibility of the DM.
We scan over $\Lambda_{eff}~{\rm or}~\Lambda_{\ell} <1500 $ TeV, $M_{\rm mess}<10^{11}$ GeV and $2.5\le\tan\beta\le 50$
and impose the following constraints:
\begin{itemize}
\item $m_{h}>110$ GeV \cite{Barate:2003sz}.\footnote{This is lower than the LEP bound $m_{h^0}>114.4$ GeV, 
justified by uncertainties involved in the Higgs mass calculation.}
\item $m_{\tilde{\chi}^{\pm}_1}>100$ GeV \cite{Abbiendi:2003sc}.
\item $m_{\rm charged~slepton}>100$ GeV \cite{Abbiendi:2003ji}.
\item $-3\times 10^{-5}<\Delta{\rm Br}(b\to s\gamma)<1.4\times10^{-4}$ \cite{Misiak:2006zs,Barberio:2008fa}.
\item $0<\Delta a_{\mu}<4\times 10^{-9}$ \cite{Davier:2009zi} .
\item $\Gamma(Z^0 \to \tilde{\chi}^0\tilde{\chi}^0)<2$ MeV \cite{:2005ema}.
\item $0.01 <\Omega h^2<0.2$ \cite{Komatsu:2008hk}. \footnote{
We have examined the finite number ($\sim 10^6$) of parameter points.
If we impose the  constraint from  WMAP DM abundance strictly, 
only tiny number of events can survive, which is statistically insufficient for the parameter search.
To enhance the number of parameter points which survive the constraints,
we loosen the condition for the DM abundance.
We expect that the allowed region in  Fig.~\ref{fig:result} does not change significantly
even if we impose severer constraint on the DM abundance.
This is because, in the region $\Omega h^2\sim 0.1$, the DM abundance depends  on the input parameters  more strongly than other physical parameters.
}
\item $\Lambda_{eff}~{\rm or}~\Lambda_{\ell} <M_{\rm mess}$.
\end{itemize}
In Fig.~\ref{fig:result}, we show the allowed region as a function of the DM mass and spin-independent cross section.
In the mGMSB, the correct DM abundance is achieved with the coannihilation effects, and 
in the non-mGMSB, focus-point like and/or $Z^0/h^{0}$ pole effects play important role,
for $m_{\rm DM}\lsim 60$ GeV.
As for $m_{\rm DM}\gsim 100~\GEV$, there are two regions.
One is the coannihilation region like mGMSB cases.
The other is  focus-point like region.
The latter region has large $\sigma_{\rm SI}$.
In the non-mGMSB cases, the colored SUSY particles have lower masses than the mGMSB ones (see Fig.~\ref{fig:mass}).
Therefore, the value of $\mu$ tends to be small, which causes large mixing of the Bino and Higgsino components.

\section{Conclusion and discussion}
In this paper, we investigated the phenomenological aspect of the GMSB with sequestered SUSY 
breaking.
The model is attractive, because the neutralino is the candidate of the DM and is free from the 
serious FCNC problems.
We especially studied the lighter DM scenario, 
taking into consideration the possibility of direct detection of the DM.
Although the recent CDMS result is still too tenuous to confirm the DM detection, 
one cannot reject the possibility of the DM signals. 
Then, the CDMS result may prefer the lighter DM.
If the light DM with $m_{\rm DM}\lsim 100$ GeV is measured in future experiments, 
minimal type GMSB is implausible, since the cross section to a nucleon of the DM is small, and 
the charged SUSY particles tend to be light, which conflict with the LEP bound.
In this paper, we discussed a modification of mGMSB to evade the current experimental bounds.

In more general gauge mediation models, it is known that the masses of the three gauginos $\tilde{g}$, $\tilde{W}$ and $\tilde{B}$
can be completely independent.
In Ref.~\cite{Meade:2008wd} it was shown that gauge mediation models in general can be parametrized by six parameters,
 $\L_{Ga}$ and $\L_{Sa}^2$ ($a=1,2,3$), by which the gaugino and sfermion masses are given by
 \beq \label{eq:gaugino-mass}
 M_a = \frac{\a_a}{4\pi} \L_{Ga}~~~(a=1,2,3), 
 \eeq
and
\beq \label{eq:scalar-mass}
m^2_{\f_i}=2\left(\frac{\a_1}{4\pi}\right)^2 C_1(i)\L_{S1}^2+2\left(\frac{\a_2}{4\pi}\right)^2
C_2(i)\L_{S2}^2+2\left(\frac{\a_3}{4\pi}\right)^2C_3(i)\L_{S3}^2.
\eeq
It was shown that all the parameter space is realizable~\cite{Carpenter:2008wi,Buican:2008ws} by constructing toy messenger models. 
Notice that the above formulae in particular show that the three gaugino masses are completely free parameters in general.
In this general framework, it is easy to evade the experimental bounds.
Although the parameter space is quite huge to investigate,
the essence of the light neutralino DM seems to be common.
It is expected that sizable mixing of the Bino and Higgsino and possibly Higgs or $Z^0$ pole effects are important.
In this case, the third family quark jets are expected to be characteristic signals just like a focus-point in the gravity mediation
\cite{Chattopadhyay:2000qa,Mercadante:2005vx,De Sanctis:2007td,Das:2007jn}, if the gluino is relatively light, compared to the masses of the squarks.
 
 The future XENON100 experiment will reach $\sigma_{\rm SI} \sim 2\times 10^{-9}~{\rm pb}$.
 In this sensitivity,  much of the parameter space with 
 $m_{\rm DM} \lsim {\cal O}(100)$ GeV can be covered.

Note added: while this work was being completed, Ref.~\cite{Ibe:2009pq} appeared. They also studied  neutralino DM models in gauge mediation, paying attention to the FCNC problems.

\vspace{5mm}

\section*{Acknowledgement}
We thank K. Hamaguchi for discussion. 
The authors are also grateful for F.~Takahashi and T.~T.~Yanagida for the collaboration 
in the early project.
This work was supported by World Premier International Center
Initiative (WPI Program), MEXT, Japan.  The work of SS and KY is supported in
part by JSPS Research Fellowships for Young Scientists.

\end{document}